\documentclass[conference,compsoc]{IEEEtran}
\IEEEoverridecommandlockouts
\pdfoutput=1
\usepackage{cite}
\usepackage{amsmath,amssymb,amsfonts}
\usepackage{algorithmic}
\usepackage{graphicx}
\usepackage{textcomp}
\usepackage{xcolor}
\usepackage{siunitx}
\usepackage{booktabs}
\usepackage{tabularx}
\usepackage{qcircuit}
\usepackage{subcaption}
\usepackage{comment}
\usepackage[colorlinks=true, hyperindex, breaklinks, linkcolor=blue, urlcolor=blue, citecolor=blue]{hyperref} 
\usepackage[capitalise]{cleveref}
\usepackage[normalem]{ulem}

\newcommand{\ket}[1]{|#1\rangle}  

\RequirePackage{ifthen}

\newboolean{ShowComments}
\setboolean{ShowComments}{true}  
\ifthenelse{\boolean{ShowComments}}%
	{
		\newcommand{\ColorComment}[3]{%
				{\colorbox{#1}{\color{white}   \textsf{\textbf{#2}}} \textcolor{#1}{#3}}}

	}%
	{
		\newcommand{\ColorComment}[3]{}

	}%

\definecolor{cocoricolor}{RGB}{200, 110, 180}
\definecolor{michalcolor}{RGB}{255,127,80}
\definecolor{naphanncolor}{RGB}{112, 51, 173}
\definecolor{porametcolor}{RGB}{198,53,39}
\definecolor{rdvcolor}{rgb}{0,0.5,0}
\definecolor{sujincolor}{rgb}{0,0,1}
\definecolor{theerapatcolor}{RGB}{47,123,245}

\def\BibTeX{{\rm B\kern-.05em{\sc i\kern-.025em b}\kern-.08em
    T\kern-.1667em\lower.7ex\hbox{E}\kern-.125emX}}
\begin{document}

\title{Boosting end-to-end entanglement fidelity in quantum repeater networks via hybridized strategies\\}

\author{
    \IEEEauthorblockN{
        Poramet Pathumsoot\IEEEauthorrefmark{1},
        Theerapat Tansuwannont\IEEEauthorrefmark{3},
        Naphan Benchasattabuse\IEEEauthorrefmark{1},
        Ryosuke Satoh\IEEEauthorrefmark{1},\\
        Michal Hajdu\v{s}ek\IEEEauthorrefmark{1}, 
        Poompong Chaiwongkhot\IEEEauthorrefmark{5}, 
        Sujin Suwanna\IEEEauthorrefmark{4}, 
        and Rodney Van Meter\IEEEauthorrefmark{2}
    }\\
    \IEEEauthorblockA{
        \IEEEauthorrefmark{1}
        \textit{Graduate School of Media and Governance, Keio University Shonan Fujisawa Campus, Kanagawa, Japan}
    }
    \IEEEauthorblockA{
        \IEEEauthorrefmark{2}\textit{Faculty of Environment and Information Studies, Keio University Shonan Fujisawa Campus, Kanagawa, Japan}
    }
    \IEEEauthorblockA{
        \IEEEauthorrefmark{3}\textit{Center for Quantum Information and Quantum Biology, Osaka University, Toyonaka, Osaka 560-0043, Japan}
    }
    \IEEEauthorblockA{
        \IEEEauthorrefmark{4}\textit{Optical and Quantum Physics Laboratory, Department of Physics, Faculty of Science,}\\\textit{Mahidol University, Bangkok 10400, Thailand}
    }
    \IEEEauthorblockA{
        \IEEEauthorrefmark{5}\textit{National Astronomical Research Institute of Thailand, Chiang Mai, 50180, Thailand} \\}
    \{poramet,whit3z,michal,rdv\}@sfc.wide.ad.jp, t.tansuwannont.qiqb@osaka-u.ac.jp, \\ poompong.ch@gmail.com, sujin.suw@mahidol.ac.th    
}

\maketitle

\begin{abstract}
Quantum networks are expected to enhance distributed quantum computing and quantum communication over long distances while providing security dependent upon physical effects rather than mathematical assumptions.
Through simulation, we show that a quantum network utilizing only entanglement purification or only quantum error correction as error management strategies cannot create Bell pairs with fidelity that exceeds the requirement for a secured quantum key distribution protocol for a broad range of hardware parameters.
We propose hybrid strategies utilizing quantum error correction on top of purification and show that they can produce Bell pairs of sufficiently high fidelity.
We identify the error parameter regime for gate and measurement errors in which these hybrid strategies are applicable.
\end{abstract}


\section{Introduction} \label{section:introduction}
\begin{figure*}[ht]
    \centering
    \includegraphics[width=0.9\textwidth]{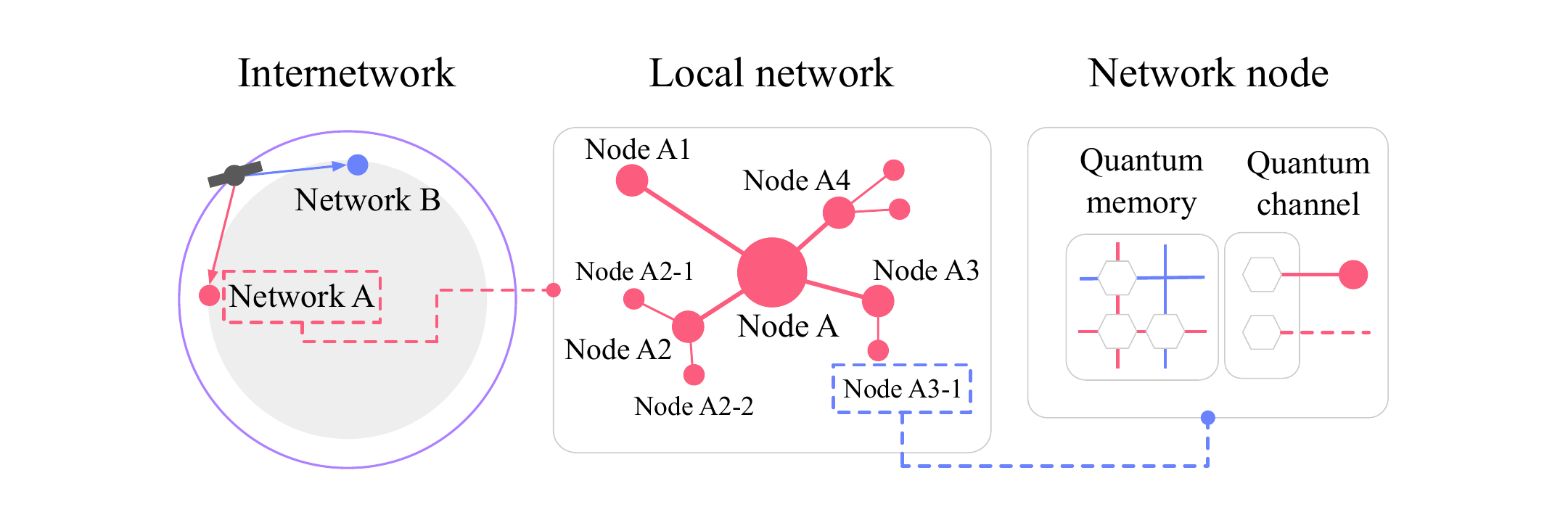}
    \caption{Concept of a worldwide quantum Internet. Inter-continental quantum communication may be realized with the aid of a network of satellites \cite{Khatri2021, Mol2023} or via SneakerNet~\cite{devitt2013quantum}, which act as long-distance Bell pair distributors for main hubs such as Node A and Node B. These hubs divide the long-distance Bell pairs with short-range entangled states generated by the local networks.}
    \label{fig:QNCopcept}
\end{figure*}

Quantum communication is at the forefront of the second quantum revolution~\cite{dowling2003quantum}.
It is expected to be instrumental in unlocking the true potential of quantum computers~\cite{arute2019quantum,zhong2020quantum,madsen2022quantum} via distributed quantum computing~\cite{cuomo2020towards,caleffi2022distributed}.
Other applications include secure quantum key distribution~\cite{ekert1991quantum,yin2017satellite,sasaki2017quantum}, improved arrays of sensing devices~\cite{gottesman2012longer,bartlett2007reference,ilo-okeke2018remote}, and secure private cloud services~\cite{fitzsimons2017private,broadbent2009universal,hayashi2018self}, which will intrinsically rely on quantum networks~\cite{vanmeter2014quantum,rfc9340}. 
Following the evolution path of the classical Internet, it is expected that quantum networks themselves will one day be connected and lead to the ultimate goal of quantum communication in the form of a global quantum internet~\cite{kimble2008quantum,wehner2018quantum,satoh2021attacking}, as illustrated in Fig.~\ref{fig:QNCopcept}.

Quantum networks utilize light as the primary information carrier between nodes of the network.
Unlike their classical counterparts, arbitrary quantum states cannot be copied faithfully~\cite{park1970concept,wootters1982single,dieks1982communication}, making classical signal amplification schemes impossible to apply \cite{chia2019phase}.
Furthermore, as quantum nodes exchange information via single photons, attenuation becomes a major obstacle to scaling of quantum networks. One approach to overcoming the exponential attenuation in fiber is to segment the entire connection between the quantum nodes using \emph{quantum repeaters}~\cite{briegel1998quantum,azuma2023quantum,hajdusek2023quantum}.
Link-level entanglement is distributed between neighboring repeater stations, which then splice them via \emph{entanglement swapping}~\cite{PhysRevLett.71.4287, pan1998experimental} into an end-to-end entangled connection.

Besides entanglement swapping, another important role of the quantum repeaters is to participate in error management.
Decoherence in quantum memories, imperfect quantum gate operations, and measurement errors all introduce unwanted \emph{operational errors}, which must be mitigated in order to ensure the fidelity remains above the threshold as required by the application.

Error management schemes can be classified into three generations~\cite{muralidharan2016optimal}.
The first generation (1G) of repeaters relies on entanglement purification schemes~\cite{briegel1998quantum,bennett1996purification,pan2001entanglement,dur2007entanglement} to detect errors on physical qubits.
The need for two-way classical communication between distant nodes in the network makes this generation of repeaters unsuitable for long-distance quantum communication.
However, due to their relatively modest hardware requirements and low resource overhead, they are expected to be the primary method of error management in early implementations of small quantum networks.
The second generation (2G) of repeaters~\cite{jiang2009quantum,fowler2010surface} uses quantum error correction (QEC) to both detect and correct errors.
This generation of repeaters avoids the issue of two-way communication and is therefore more suitable for long-distance quantum communication, albeit at the price of stricter hardware requirements and larger resource overhead.
The third generation (3G) of quantum repeaters~\cite{muralidharan2014ultrafast,munro2012quantum} relies on quantum error correction to correct both photon loss and operational errors and does not require entanglement swapping or pre-shared long-distance entanglement as a means of communication.
Thus, 3G repeaters place far more strict requirements on loss and operational fidelity and place more complex demands on the hardware than 2G repeaters.

The performance of individual repeater generations has been analyzed and contrasted in previous reports \cite{muralidharan2016optimal,dur1999quantum,fujii2009entanglement,jansen2022enumerating}.
Apart from the generation of the repeater network, the network performance also depends on the entanglement distribution policy \cite{khatri2021policies}.
In this work, we present a comprehensive study of a number of entanglement distribution strategies and compare their performance under realistic noise conditions in terms of end-to-end fidelity and throughput. We show by simulation that using either purification or QEC alone is not enough to produce end-to-end Bell pairs with sufficiently high fidelity above a certain noise level. To solve this problem, 
instead of deploying QEC and purification techniques separately, we utilize them together by introducing a hybrid generation (HG), which we refer to as \emph{purified encoding} (PE).
PE first increases the fidelity of physical Bell pairs by using purification and then encodes a logical Bell pair. We investigate two variations of this hybrid strategy. The first one is referred to as as HG-PE in which we perform PE immediately after producing a physical Bell pair at the link level. In the second strategy, denoted by E2E-HG-PE, we employ PE only at the end nodes. In these strategies, we identify the error parameter regime, e.g. quantum gate errors and measurement errors, where HG outperforms strategies based purely on 1G or 2G.

\section{Error Management in Quantum Repeater Networks} \label{section:fromTheoryToSimulation}

In this section, we briefly introduce the essential elements used as building blocks in a quantum network. We then discuss the error models we used in evaluating quantum communication strategies. To that end, we would like to introduce entanglement distribution strategies used in our discussion. 

\begin{figure*}[ht]
    \centering
    \includegraphics[width=\textwidth]{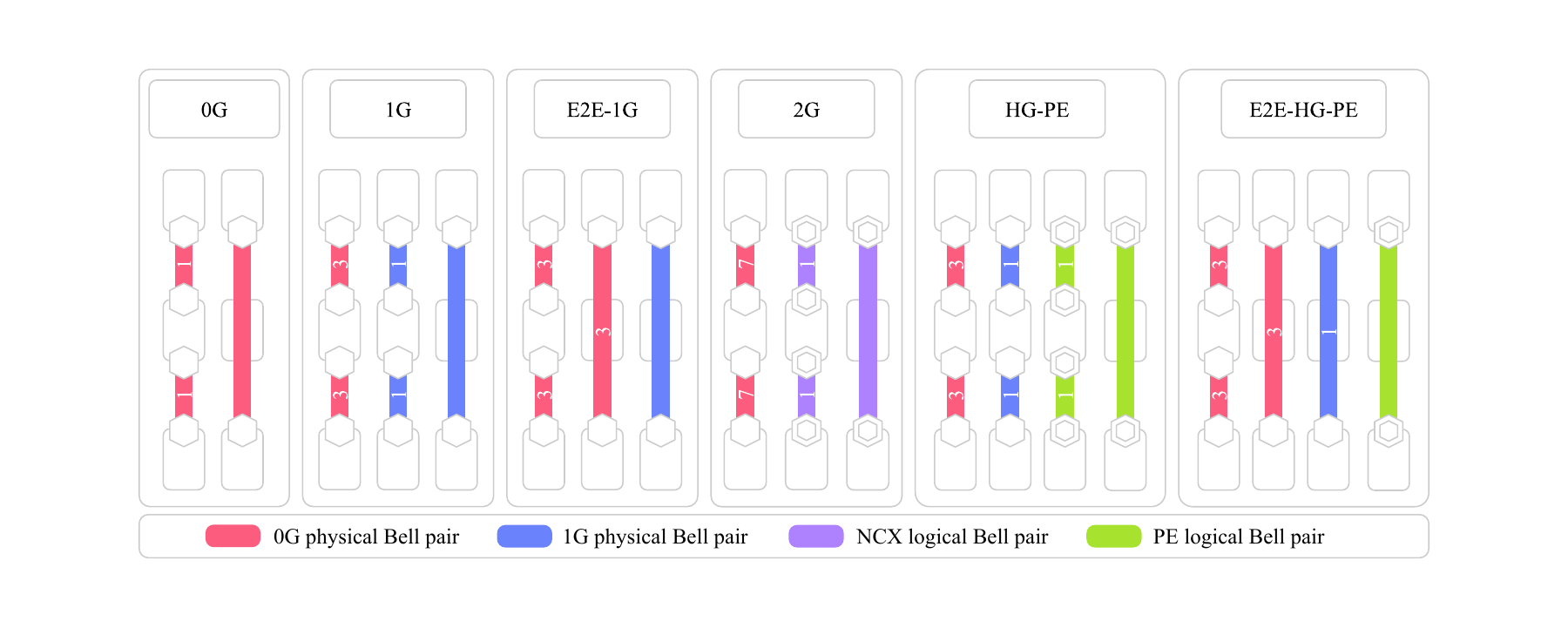}
    \caption{Basic and hybrid entanglement distribution strategies. Each strategy shows Bell pairs drawn vertically, with end nodes at the top and bottom and repeaters (if any) in between. Within each strategy, actions are shown left to right, beginning with raw link-level Bell pairs and ending with the high-fidelity end-to-end Bell pair. The numbers on the colored links indicate the number of Bell pairs from the previous round consumed in the current round.
    }
    \label{fig:strategies}
\end{figure*}

As mentioned in \cref{section:introduction}, current quantum network architectures are divided into three generations. Here, we considered 0G, 1G and 2G, as 3G is less likely to be achievable in the near future. The overview scheme of the strategies considered in our work are illustrated in \cref{fig:strategies}.

\textbf{[0G] strategy:}
Considering the 1G and 2G networks, the most basic building blocks of each strategy are entanglement swapping, and link-level generation of Bell pairs.
This link-level entanglement generation is a process where each pair of neighboring nodes continuously (or upon request) produces Bell pairs for consumption.
Entanglement swapping is a process that takes two Bell pairs from nodes (A, B) and nodes (B, C) and produces one Bell pair for node (A, C) as the output. This process is not limited to Bell pairs from two nodes that are adjacent to each other. This leads us to the first strategy which we refer to as [\textbf{0G}]. In this strategy, only entanglement swapping and link-level generation of physical Bell pair are used. An overall strategy is to perform entanglement swapping until the Bell pairs are produced for both end-nodes. The only complication of this level is the order of entanglement swapping, where the linear path for both end-nodes does not have only one node in between. There are two general approaches for the problem. The first approach is to perform entanglement swapping all at once at each intermediate node.
The measurement results for Pauli correction required for each swapping will be sent to both end nodes, where only one side need to do cumulative correction, while sending results to another node serves as a signal to notify the entanglement swapping result.
The second approach is ordered entanglement swapping where for swapping, the Bell basis measurement and correction need to be finished before proceeding to next swapping.

Consider an example of four nodes with linear topology, which produce Bell pairs at nodes 1 and 4. We choose to perform entanglement swapping at node 2 first to create a Bell pair between nodes 1 and 3.
We then proceed to entanglement swap at node 3.
Contrasted to simultaneous entanglement swapping that nodes 2 and 3 could perform at their convenience, the ordered approach requires the qubits to be locked for much longer.
This means that the qubits cannot participate in other processes at that time. Consequently, node 3 has to wait until its qubit is released.
This does not imply that we cannot do better as we could optimize this approach by performing entanglement swapping when possible and without pre-determined order.
Consider a longer linear path.
There exist nodes in which their qubits are not locked by other processes, they can perform entanglement swapping in parallel with other nodes in the path. 

This optimization, however, does not give a speed comparable to the simultaneous approach, but it gives an opportunity to enhance quality of Bell pairs produced after and before each entanglement swapping using the enhance method introduced by 1G and 2G, which otherwise is not possible with simultaneous approach. 
We choose to perform ordered entanglement swapping for all of the strategies including [\textbf{0G}] for the sake of simplicity.

\textbf{[1G] and [E2E-1G] strategies:}
For 1G, entanglement purification is an addition process on top of the primitive processes of [\textbf{0G}]. Purification takes multiple Bell pairs as an input and produces a Bell pair with higher fidelity. The relation of input and output fidelity depends on the choice of a purification protocol. In the simplest case, X-purification and Z-purification take two Bell pairs, and output one Bell pair. Even with the low rate of resource consumption, as the name implied, they could only correct for bit-flip (Pauli X) or phase-flip (Pauli Z) errors, which is unlikely in real-world application to have only bit-flip or phase-flip errors. Even if the higher fidelity Bell pair is produced, the fidelity is not unity; in other words, the quality of improved Bell pair is not perfect. Purification could be perform on top of each other, to produce even higher fidelity Bell pair with cost of resource consumption.
Furthermore, the purification is a probabilistic process; thus, if the fidelity of input Bell pair is low, the probability of successful purification will decrease, and so delay the operation even more.

There are many possible strategies that could be performed using only purification. In this work, we consider two types of strategies based on purification.
The first one performs purification once after link-level entanglement generation, referred to as [\textbf{1G}].
The second strategy executes purification on the end node Bell pairs before the resource consumption, which we refer to as [\textbf{E2E-1G}]. The motivation of the latter strategy is to utilize the speed of [\textbf{0G}] and then enhance the quality of fresh Bell pair for later application with low latency. The purification protocol that we use in this work is the Ss-Dp protocol \cite{fujii2009entanglement,matsuo2019quantum, Matsuo2019} where three Bell pairs are required for this protocol.

\textbf{[2G] strategy:}
For 2G, quantum error correction (QEC) is used to improve fidelity of the Bell pair. 
Similar to classical error correction in which a logical bit is encoded using many physical bits, QEC encodes a logical qubit using many physical qubits. 
To identify the type of errors and the physical qubits on which the errors occurred, operator measurements involving auxiliary qubits called syndrome measurements are performed. Afterwards, the error correction based on the error syndrome (the measurement results) can be done. As QEC moves computation to the logical level, the logical error rate can be suppressed if the physical error rate is below some threshold value. That is, decoherence such as decay could be alleviated. We can view QEC as a possible way to preserve quantum state for an extended period of time. However, since QEC requires more resources with high fidelity, the throughput of the network is reduced, making 2G more difficult to realize than 1G. In our work, we perform QEC using the Steane code \cite{Steane1996}, a quantum error correcting code that encodes one logical qubit into seven physical qubits. We first prepare two logical qubits in states $\ket{0}$ and $\ket{+}$, then create a logical Bell state by applying transversal non-local CNOT (NCX) gates, which effectively are a logical CNOT gate between logical qubits from different nodes. 

\textbf{[HG-PE] and [E2E-HG-PE] strategies:}
Having discussed the advantages/disadvantages of each generation individually, we now consider the possibility of utilizing them together. In the first variation, we wish to reduce the resources required to create link-level logical Bell pairs. Instead of NCX gates, we consider using high fidelity physical Bell pairs delivered by purification shared between nodes as an input for QEC. This method directly produces a logical Bell pair. We refer to this strategy as [\textbf{HG-PE}]. A variation of this strategy, where we perform the purification and encoding at the end nodes is denoted by [\textbf{E2E-HG-PE}].
The discussion of theses two variations will be focused in \cref{sec:hybrid_error_management}.

\section{Hybrid Error Management} \label{sec:hybrid_error_management}

\begin{table*}[ht]
\begin{center}
\begin{tabular}{@{}lllllll@{}}
\toprule
\textbf{Requirements}                                                               & [\textbf{0G}]      & [\textbf{1G}]      & [\textbf{E2E-1G}]  & [\textbf{2G}]       & [\textbf{HG-PE}]    & [\textbf{E2E-HG-PE}] \\ \midrule
Generation requirement for intermediate-nodes                              & 0G        & 1G        & 0G        & 0G, 2G     & 0G, 1G, 2G & 0G          \\ \midrule
Generation requirement for end-nodes                                       & 0G        & 1G        & 1G        & 0G, 2G     & 0G, 1G, 2G & 2G          \\ \midrule
Minimum number of physical qubits per QNIC \\for intermediate-nodes (E + I) & 1 (1 + 0) & 3 (3 + 0) & 1 (1 + 0) & 14 (7 + 7) & 9 (3 + 6)  & 1 (1 + 0)   \\ \midrule
Minimum number of physical qubits per QNIC \\for end-nodes (E + I)          & 1 (1 + 0) & 3 (3 + 0) & 3 (3 + 0) & 14 (7 + 7) & 9 (3 + 6)  & 9 (3 + 6)   \\ \midrule
E2E resource type                                                          & Physical  & Physical  & Physical  & Logical    & Logical    & Logical     \\ \bottomrule
\end{tabular}
\caption{The table summarizes requirements for each strategy in terms of resource consumption and technological requirements of each node. E and I in (E + I) stand for External and Internal qubits. External qubits refer to qubits that connect to a quantum communication channel between nodes. And internal qubits are qubits mainly used within the node, and not required to have direct connection to the communication channel.}
\label{table:strategies_requirement}
\end{center}
\end{table*}

Let us discuss in-depth motivations, use cases and factors one might need to consider when deciding which entanglement distribution strategies to deploy.
It is important to consider resource and technological requirements for quantum nodes within the network in order to reduce the infrastructure cost. 

We consider simple architecture abstraction, where every quantum node with quantum memories to store physical qubits.
It also has the ability to manipulate the qubits with at least a set of Clifford gates, perform measurements in the Pauli basis, and generating link-level Bell pairs with neighboring nodes.
We refer to this equipment as \textit{quantum network interface card} (QNIC)~\cite{Matsuo2019,vanmeter2022quantum}.
For each connection between two distant nodes, each node will possess one QNIC. For example, in a linear network, intermediate nodes will have total of two QNICs, and end-nodes will have one QNIC. We assume that all qubits within the nodes have all-to-all connectivity for simplicity regardless of QNIC that qubit contained within for the discussion and simulation. However, this is less likely to be the case in the realistic situations, depending on the type of physical qubits realization. However, it also serves as a point of consideration for the case that number of qubits is an important factor such that highly connected architecture is likely to be achievable for devices with smaller number of qubits. The delay and operations needed for distant qubits to operate CNOT gates together will also open a window for error to happen.

As investigated in Ref.~\cite{PhysRevA.93.042338}, the use of purification before encoding logical qubit to produce logical Bell pair between nodes and difference QEC codes is not surprising.
In our case, we consider the use of purification for encoding as a method to reduce the cost of preparing high fidelity logical Bell pairs. Compared to non-local CNOT gate used in [\textbf{2G}], this method is more straightforward and simpler to implement. 

Important consideration when choosing a suitable network architecture is the purpose of the network.
For example, quantum network can be used for distributed quantum computing or for quantum key distribution.
Both of these applications come with different requirements that the architecture must satisfy.
Two crucial questions that must be answered are ``How should one distribute Bell pairs?'', and ``What type of Bell pairs are required?''
For key distribution, it may be sufficient to distribute physical Bell pairs in the early days of development.
For distributed computation, logical Bell pairs high-fidelity requirements may call for the use of logical Bell pairs. 

In the case of logical Bell pairs, hybrid strategies are a natural solution both in terms of quality and throughput of the network.
We will demonstrate below that even if the desired Bell pairs are physical, hybrid strategies outperform pure 0G and 1G in terms of fidelity. 

\cref{table:strategies_requirement} shows the requirements for intermediate nodes and end nodes.
The requirements for the intermediate nodes are less demanding compared to the requirements for the end nodes.
The end nodes need to perform more processing on the requested resources.
The requirements are further divided in terms of the number of qubits required apart from the generation of the node itself.
Number of qubits required per request does matter as the lower the number the more requests can be handled simultaneously by the node, which subsequently results in higher throughput.

[\textbf{2G}] and [\textbf{HG-PE}] strategies are appropriate if the resource needs to be store in intermediate nodes for some time before being being processed further.
These strategies produce encoded link-level Bell pairs before entanglement swapping in order to extend the lifetime of the resource.
This is in contrast to [\textbf{E2E-HG-PE}], which encodes logical Bell pairs at the end nodes only.
[\textbf{HG-PE}] requires the intermediate nodes to be equipped with more memory qubits compared to 1G strategies.
However, we expect that that 2G capable nodes will have no difficulty running 1G protocols.
Additionally, [\textbf{HG-PE}] requires less qubits than the pure [\textbf{2G}] strategy.
The [\textbf{E2E-HG-PE}] strategy will be the preferred choice if low latency in resource consumption at the end nodes is desired since the qubit consumption per QNIC for intermediate node is as low as [\textbf{0G}].
The resource can be continuously produced, and the accumulated resources can be stored for a long period of time. 

Hybrid strategies are likely to be achievable in the near future compare to 3G as they use relatively simple QEC and small number of purifications. Even with simple QEC as Steane code, they allow us to do classical error correction on measurement results on top of quantum error correction. Steane code is not the only code that classical error correction could be done after the measurement, but it is enough for us to demonstrate the usefulness of it as a representative of its own class. Further numerical analysis which includes errors in processing and number of intermediate nodes are present in the \cref{section:result}.

\section{Simulation Setting and Parameters} \label{section:setting}

In this section, we motivate the configuration of our simulation before discussing the numerical analysis of entanglement distribution strategies.
The strategy to establish long-range entanglement depends strongly on the noise regime as well as the demands of the intended application.
These requirements are often formulated in terms of threshold fidelity \cite{Perseguers2010}.
For example, a QKD application that is robust against the faked Bell state attack requires an end-to-end fidelity of at least 0.83~\cite{Sajeed2019BrightlightDC}.
We use this threshold fidelity as a convenient goal for the considered strategies and show that the purely 1G or purely 2G distribution strategies are unable to reach it under our simulation conditions.
On the other hand, our new hybrid strategies are capable of surpassing this threshold fidelity for some noise regimes which we identify.

We now provide details of all the fixed parameters that we use in our simulations.
\begin{enumerate}
    \item We consider a linear chain of repeaters with the distance between the two end nodes being fixed at 100 \si{\km}. The link lengths are all equal but the number of repeaters can be varied.
    \item The speed of light in fiber is assumed to be constant at \num{300000} \si{\km/\s}, although, it does not represent a realistic speed of light in fibre. This choice stems from the fact that the distances between two adjacent nodes in our linear path are equal. The figure serves as a constant delay of photon travelling between nodes and are the same for all path.
    \item The depolarizing probability is fixed at $p_{\text{depo}}=0.025$. It should be noted that the entanglement generation over a distance of 50 km using $^{40}$Ca$^{+}$ ion in Ref.\cite{Krutyanskiy2019} reported a Bell pair of fidelity of 0.86 (simulation) and 0.86 $\pm$ 0.03 (experiment), where the simulation accounted for the effect of measured background counts only. Compared to our work, including only depolarizing channel applied directly on the ideal Bell pair, setting $p_{\text{depo}}$ to be approximately 0.0736 results in the output fidelity of approximately 0.86.
    \item The memory lifetime of $\tau = 10 \text{ ms}$. This generous lifetime is typical in a number of physical systems, particularly in ion traps~\cite{Wang2021} and NV centers in diamond~\cite{pompili2021realization}. 
    \item We assumed the effective loss rate to be \num{0.30} \si{dB/km}.
    This is a conservative value; optical fibers with substantially lower attenuation have been used in experiments.
    This value is sufficient to affect the quality of qubits that suffer from a long waiting time.
\end{enumerate}

The depolarizing channels is applied to link-level Bell pairs with probability $p_{\text{depo}}$ immediately after their creation.
Gate errors are applied every time a quantum gate is applied with probability $\lambda_{\text{gate}}$.
This involves operations such as entanglement swapping, NCX, and syndrome measurement.
Furthermore, immediately before a measurement, we apply a memory error in the form of a depolarizing channel, with strength depending on the qubit lifetime $\tau$, and the time elapsed since initialization of the qubit.
In order to simulate measurement errors, bit flip error is applied to the measurement result with probability $p_{\text{meas}}$.

The above parameters have typical, or worse than typical, values compared to the experimental ones reported in Refs.~\cite{Hofmann2005, Liao2017, Chen1989, Valentini2021AnalysisOP, Abasifard2023}, and are kept fixed in the strategy evaluation. Variable parameters for optimization are the gate error parameter $\lambda_{\text{gate}} \in \{0.0000,0.0005,0.0010,0.0015,0.0020\}$, the measurement error $p_{\text{meas}} \in \{0.0000,0.0025,0.0050,0.0075,0.0100\}$, and the number of hops $h_s \in \{2, 4, 8\}$.

\section{Results}\label{section:result}

In this section, we address the regime where hybrid strategies can reach fidelity thresholds that pure strategies do not. 

\begin{figure*}[t]
    \centering
    \includegraphics[width=0.9\textwidth]{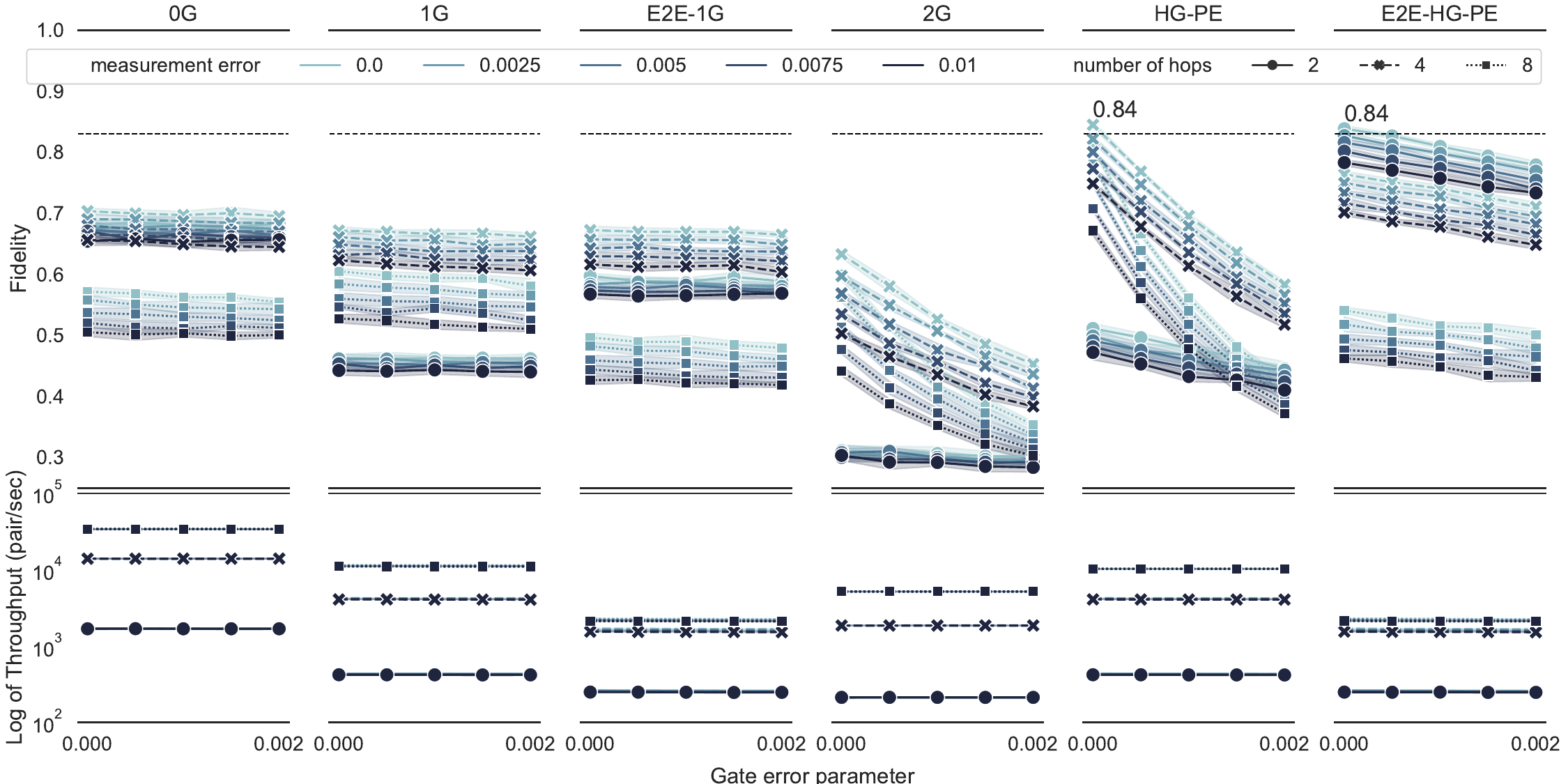}
    \caption{The upper plots show the fidelity of the end-to-end Bell pair yielded from each strategy with varying gate error parameter $\lambda_{\text{gate}} \in \{0.0000,0.0005,0.0010,0.0015,0.0020\}$, measurement error $m_e \in \{0.0000,0.0025,0.0050,0.0075,0.0100\}$ and the number of hops $h_s \in \{2, 4, 8\}$. The dashed line is the reference fidelity of value 0.83. The lower plots show the throughput of each strategy on a logarithmic scale.}
    \label{fig:fidelity}
\end{figure*}

~\cref{fig:fidelity} summarizes our results both in terms of fidelity and throughout. The horizontal dashed lines indicate the threshold fidelity of 0.83 that we aim to reach. The results clearly show that both variants of hybrid strategies, namely  [\textbf{HG-PE}] and [\textbf{E2E-HG-PE}], are the only two strategies that arecapable of reaching this threshold, even with a very optimistic regime of near zero gate and measurement errors. However, we emphasize that this result holds in the presence of fixed depolarizing errors, memory errors, and photon loss at certain values as indicated in \cref{section:setting}. Further improvement of these noise parameters would likely allow for more room for gate and measurement errors. 

Keeping the probability of depolarizing errors rate $p_{\text{depo}}$ fixed at 2.5\%, the fidelity of Bell pairs generated between adjacent nodes is independent of the distance. However, for some strategies, the fidelity does not always decrease as the number of hops increases. This seemingly counter-intuitive observation is a direct consequence of the fact that the total distance between two end nodes is fixed. Increasing the number of hops therefore shortens the link length. Combined with finite memory lifetime, this results in the observed behavior indicating that simply increasing the number of hops can lead to higher end-to-end fidelity.

From \cref{fig:fidelity}, we note that simply distributing the Bell pairs fast is not sufficient as [\textbf{0G}] also fails to reach the threshold fidelity of 0.83. Hence, a noise suppression technique is warranted. However, it is evident that [\textbf{1G}] also could not produce Bell pairs that meet the fidelity threshold. Even its extension, [\textbf{E2E-1G}], suffers from short memory time such that the purification could not offer any advantage over [\textbf{0G}]. Even worse, [\textbf{2G}] requires more physical resources, effectively prolonging the waiting time and yields Bell pairs of fidelity lower than other strategies. From these results, we regard short memory lifetime as a main problem, consistent with a recent report \cite{Mol2023}. We also remark that the sensitivity to the gate errors could be separately considered in two groups: (i) the strategies that use up to 1G technology, namely [\textbf{0G}], [\textbf{1G}], and [\textbf{E2E-1G}]; and (ii) the strategies that use up to 2G technology, namely [\textbf{2G}], [\textbf{HG-PE}], and [\textbf{E2E-HG-PE}]. The first group is less sensitive to the gate errors than the second group, as evident from \cref{fig:fidelity}. This is because 2G requires many more qubits and high number of gates for operations in error detection and correction. Also, measurement errors do not cause drastic change as gate errors. In addition, one observation that both hybrid strategies differ from each other is that [\textbf{E2E-HG-PE}] is less sensitive to gate errors since QEC is applied only at the end-to-end level Bell pair. 

\section{Discussion} \label{section:discussion}

Our results show that the failure to realize high fidelity quantum communication is a result from short memory lifetime. 
It is generally not a good strategy to simply increase the number of hops along the path to increase the end-to-end Bell pair production rate as it unavoidably induces more errors in the system due to imperfections of the applied operations.
A lower number of quantum repeaters also leads to the decoherence of qubits due to increased waiting times during which quantum memories are locked.

Our proposed solution is to use a hybrid strategy. \cref{fig:fidelity} shows that in order to deliver high fidelity Bell pairs to the application, it is desirable to distribute physical Bell pairs with 0G between end nodes followed by purification to increase their fidelity, before finally encoding them as logical Bell pairs.

End-to-end Bell pairs can be generated by the quantum network in advance and then consumed on demand by the application.
Using purification and QEC helps preserve the fidelity of the end-to-end Bell pair until it is needed by the application.

It would be interesting to consider [\textbf{0G}] to distribute Bell pairs for encoding at the destination nodes, or other purification schemes \cite{fujii2009entanglement, dur2007epa}.
As our memory error model is quite simple, more advanced models might reveal other interesting behavior of the network.

\section*{Acknowledgement}
This research has received funding support from  Mahidol University (Fundamental Fund FF-093/2567: fiscal year 2024 by National Science Research and Innovation Fund (NSRF)) and the NSRF via the Program Management Unit for Human Resources \& Institutional Development, Research and Innovation [grant number B05F650024]. 
The work of NB, MH, and RDV was supported by JST [Moonshot R\&D Program] Grant Number [JPMJMS226C].

\bibliographystyle{IEEEtran}
\bibliography{references_2}

\end{document}